\documentclass[12pt]{article}
\input{epsf.sty}

\def\1{{\mathchoice {1\mskip-4mu\mathrm l}
{1\mskip-4mu\mathrm l} {1\mskip-4.5mu\mathrm l}
{1\mskip-5mu\mathrm l}}}

\usepackage{hyperref}
\usepackage{amsmath, amssymb} 
\usepackage{amsmath, amssymb} 
\newenvironment{remark}[1][Remark:]{\begin{trivlist}
\item[\hskip \labelsep {\bfseries #1}]}{\end{trivlist}}

\def\N{{\mathbb N}}
\def\Z{{\mathbb Z}}
\def\R{{\mathbb R}}

\begin{document}
\title{An ultrametric state space with  a dense discrete overlap distribution: \\
Paperfolding sequences} 
\author{
 Aernout C. D. van Enter        \\[-1mm]
  {\normalsize\it Johann Bernoulli institute}   \\[-1.5mm]
  {\normalsize\it Rijksuniversiteit Groningen}         \\[-1.5mm]
  {\normalsize\it Nijenborgh 9}                \\[-1.5mm]
  {\normalsize\it 9747 AG Groningen}           \\[-1.5mm]
  {\normalsize\it THE NETHERLANDS}             \\[-1mm]
  {\normalsize\tt A.C.D.van.Enter@.rug.nl}        \\[-1mm]
  {\normalsize Ellis de Groote}               \\[-1mm]
  {\normalsize\it De Vonderkampen 116}   \\[-1.5mm]
  {\normalsize\it 9411RG, Beilen}   \\[-1.5mm]
  {\normalsize\it THE NETHERLANDS}             \\[-1mm]
  {\normalsize\tt ellisdegroote@hotmail.com}  \\[-1mm]} 
{\protect\makebox[5in]{\quad}}
\pagenumbering{arabic}
\maketitle
\baselineskip=14pt
\noindent {\bf Abstract.} 
We compute the Parisi overlap distribution for  paperfolding sequences. 
It turns out to be discrete, and to live on the dyadic rationals. Hence it is  
a pure point measure whose support is the full  interval $[-1,+1]$. 
The space of paperfolding sequences has an ultrametric structure. 
Our example  provides an illustration of  some  properties which were suggested 
to occur for  pure states in spin glass models. 

\newtheorem{theorem}{Theorem}          
\newtheorem{lemma}[theorem]{Lemma}              
\newtheorem{proposition}[theorem]{Proposition}
\newtheorem{corollary}[theorem]{Corollary}
\newtheorem{definition}[theorem]{Definition}
\newtheorem{conjecture}[theorem]{Conjecture}
\newtheorem{claim}[theorem]{Claim}
\newtheorem{observation}[theorem]{Observation}
\def\proof{\par\noindent{\it Proof.\ }}
\def\reff#1{(\ref{#1})}

\let\zed=\bbbz 
\let\szed=\bbbz 
\let\IR=\bbbr 
\let\R=\bbbr 
\let\sIR=\bbbr 
\let\IN=\bbbn 
\let\IC=\bbbc 

\def\nl{\medskip\par\noindent}

\def\scrb{{\cal B}}
\def\scrg{{\cal G}}
\def\scrf{{\cal F}}
\def\scrl{{\cal L}}
\def\scrr{{\cal R}}
\def\scrt{{\cal T}}
\def\pfin{{\cal S}}
\def\prob{M_{+1}}
\def\cql{C_{\rm ql}}
\def\bydef{\stackrel{\rm def}{=}}   
\def\qed{\hbox{\hskip 1cm\vrule width6pt height7pt depth1pt \hskip1pt}\bigskip}
\def\remark{\medskip\par\noindent{\bf Remark:}}
\def\remarks{\medskip\par\noindent{\bf Remarks:}}
\def\example{\medskip\par\noindent{\bf Example:}}
\def\examples{\medskip\par\noindent{\bf Examples:}}
\def\nonexamples{\medskip\par\noindent{\bf Non-examples:}}

\newenvironment{scarray}{
          \textfont0=\scriptfont0
          \scriptfont0=\scriptscriptfont0
          \textfont1=\scriptfont1
          \scriptfont1=\scriptscriptfont1
          \textfont2=\scriptfont2
          \scriptfont2=\scriptscriptfont2
          \textfont3=\scriptfont3
          \scriptfont3=\scriptscriptfont3
        \renewcommand{\arraystretch}{0.7}
        \begin{array}{c}}{\end{array}}

\def\wspec{w'_{\rm special}}
\def\mup{\widehat\mu^+}
\def\mupm{\widehat\mu^{+|-_\Lambda}}
\def\pip{\widehat\pi^+}
\def\pipm{\widehat\pi^{+|-_\Lambda}}
\def\ind{{\rm I}}
\def\const{{\rm const}}

\bibliographystyle{plain}

\section{Introduction}

\bigskip

In \cite{EHM} the study of the Parisi overlap distribution
for various classes of non-periodically  ordered sequences was undertaken.
\newline
The considered  sequences were members of $\{-,+\}^{\Z}$ and their orbit 
closure forms typically 
a uniquely ergodic system, with a unique shift-invariant measure $\mu$. 
This measure then can be a ground state for a translation-invariant interaction 
which one can construct \cite{Aub1,rad,Rad1}, and the individual 
sequences will be the pure (extremal) ground states.
\newline 
Their overlap distribution gives the behaviour under the product measure of 
$\mu$ with itself, describing a two-replica system, written as  
$\mu \times \mu'$ of the overlap between two randomly 
(from this product measure) chosen, bi-infinite sequences $\sigma$ and 
$\sigma'$:
 
\begin{equation}
q_{\sigma \sigma'}=lim \frac{1}{N} \sum_{i=1,...N} \sigma_i \sigma'_{i}. 
\end{equation} 

Note that the overlap $q_{\sigma \sigma'}$ between two sequences is directly 
related to their Hamming distance $d_H(\sigma, \sigma')= \frac{1-q_{\sigma \sigma'}}{2}$. 

\smallskip

If we take two random sequences, each chosen according to the same 
shift-invariant measure, with probability one the above limit will exist, as 
follows from the ergodic theorem.
The product measure of two ergodic measures, however, although shift-invariant, 
in general will not be ergodic, thus one may obtain different values for the  
overlap with positive probability.
 
A simple example illustrating this is the symmetric measure which give equal 
weights to  the two alternating sequences:

\begin{equation} 
\mu_{alt}(\sigma) = \frac{1}{2}(\delta(+-..., \sigma) + \delta(-+..., \sigma))
\end{equation}

This measure is  ergodic under the shift transformation, but the product measure
of $\mu_{alt}$ with itself is not. For the overlaps it holds that with 
probability $\frac{1}{2}$ the same sequence is chosen (overlap one) and with 
probability  $\frac{1}{2}$ two different sequences are chosen 
(overlap minus one). Thus in this simple example 

\begin{equation} 
p(q) = \frac{1}{2}(\delta(q,1) + \delta(q,-1))
\end{equation}

In more general situations the  overlap can take any value in the interval 
$[-1,+1]$, and the induced measure, which is the overlap distribution, 
describable by a probability distribution function
on this interval, can be discrete ( that is, it is a pure point measure)
or nondiscrete, and  thus have continuous components.

In particular, it was found in \cite{EHM} that whenever the atomic 
("diffraction")  spectrum of the unique translation-invariant measure on 
the sequences does not contain a pure point component, the 
overlap distribution is trivial; this applies in particular to all weakly 
mixing systems (in  which case their product measures are ergodic), 
as well as to the (Prouhet-(Thue-))Morse system. 

On the other hand for the Fibonacci sequences the overlap distribution 
was found to contain a(n absolutely)  continuous part. 
In fact, in explicit form it is \cite{EHM} 
\begin{equation}
p(q)dq = 
(2 \gamma -1) \delta(q, 1-4(1-\gamma))+ \frac{1}{2}1_{[1-4(1- \gamma),1]}(q)dq
\end{equation}
where $\gamma= \frac{2}{1+\sqrt 5}$. 

We notice here that 
the same argument which was used in \cite{EHM}  applies to more general  
Sturmian sequences, which thus also have a continuous part in their overlap 
distribution. The reason is that these are  also rotation sequences, 
see e.g.  \cite{LW}. They also have recently been studied as ground states of 
some explicit interaction functions on lattice systems in \cite{ADJR}). 

Moreover,  it was found that the Toeplitz (or period-doubling) 
sequences give rise to an overlap 
distribution which is concentrated on a countable number of values, 
with $1$ as its only limit point. Also, the set of these Toeplitz sequences 
has a tree (ultrametric) structure, and the overlap distribution 
could be obtained via this structure.

Here we show that the paperfolding sequences (see \cite{BaMoRicSi,DekMenPoo} 
and references mentioned there), which also display such an  
ultrametric structure, give rise to an overlap distribution living on a 
{\em dense} countable 
number of points. As such the paperfolding system, which, in contrast to the  
Toeplitz system, is symmetric under the spin-flip (plus-minus) symmetry, 
is even closer to what is expected in Parisi's replica symmetry breaking 
theory of the Sherrington-Kirkpatrick model \cite{MPV, PT}.
In Parisi's theory for the SK model, this countable overlap distribution 
is supposed to show up for a fixed disorder, while averaging over the disorder 
distribution is supposed to result in a continuous piece in the overlap 
distribution. Often the countable overlap distribution is interpreted as being 
related to or implying a countable number of pure states; however there 
seems not to be much justification for this \cite{Bol}. 

\smallskip
Conceptually, our example illustrates a number of points:
\smallskip

\noindent
Newman and Stein \cite{NS1,NS2,NS3,NS4,NS5, N} have shown that self-averaging 
arguments, based on the use
of the ergodic theorem, apply to many finite-dimensional. disordered 
(spin-glass) models (although  not to the equivalent-neighbour 
Sherrington-Kirkpatrick model). 
In particular they find that for systems with a spatial structure, 
in contrast to the Parisi theory's predictions for the
Sherrington-Kirkpatrick model  \cite{MPV,PT}, overlap quantities cannot 
depend on the 
realization of the disorder parameter, that is, they are "self-averaging". 
For example, for Edwards-Anderson-type  spin-glass models there is no 
difference between  a typical (Parisi SK-prediction: supported on countably many values, dense)
and an averaged (Parisi SK-prediction: continuous) overlap distribution.

\bigskip

As mentioned before, for many nonperiodic systems, the individual 
(nonperiodic) sequences 
are ground states for some translation invariant deterministic interaction.
Thus they are often used as models for non-periodic (quasi-)crystals, 
\cite{Aub1,rad,Rad1, GERM,Mi1,Rue1,SBGC,Sen}. It therefore becomes natural to 
consider various questions naturally arising in statistical mechanics 
about such systems, whether about their spectral properties 
\cite{ BaHo,EMw, LeeMo, LeeMoSo} 
or, as was done  in \cite{EHM,E} and also here, about their overlap distributions.
Typically, there are uncountably many such ground states, and as mentioned 
before, they can  have nontrivial overlap distributions; 
somewhat surprisingly even without the presence of any disorder. 

It turns out that:
\begin{itemize}
 
\item There is no general connection between the number of pure states -uncountable-
and the number of possible overlap values --which can be one, countably or 
uncountably many--.

\item Triviality,  non-triviality, and ultrametricity of the Parisi overlap 
distribution all can be realized  without any role for the disorder.

\item A dense discrete overlap distribution, with a hierarchical structure, 
can be realised (by the paperfolding system).

\end{itemize}
\section{Main Result}

For background  on  the paperfolding system we refer to \cite{BaMoRicSi, DekMenPoo, AS}, and references mentioned there.
 
\smallskip

\noindent
A paperfolding sequence can be constructed as follows: 
\newline
In step 1, we choose $k_1$ to be a number from the set $\{1,2,3,4 \}$.
\newline
Now we fill all sites of the form $k_1+4 \Z$  with pluses, and all sites 
of the form $k_1+2+4 \Z$ with minuses.
\newline
Thus we have the period-4 structure covering half the sites:  
\newline
$  .....\   +.-.+.-.+.-.+.-. \ ...$
\newline
In step 2, we again choose $k_2$ to be a number from the set $\{1,2,3,4 \}$.
\newline
Now we occupy all sites of the form $k_1+1+ 2k_2 + 8\Z$ with pluses and all 
sites $k_1+5+ 2k_2 + 8 \Z$ with minuses.

We repeat this step and in the $m$th step we again occupy one quarter of the 
remaining empty sites  with pluses, periodically with period $2 \times 2^m$, 
and the quarter of the sites at distance $2^m$ from the pluses we occupy with 
minuses,  periodically with the same period. Thus in each step one half of the 
remaining sites are filled.  

\smallskip

After making a choice for $k_m$, for all $m \in \N$, all sites will be filled, 
and we have a paperfolding sequence.
Notice that in every step the uncountable set of sequences is divided into 
four times as many  subsets as in the previous step, which generates the 
hierarchical (ultrametric) structure of the uncountable  set of sequences. 
 

With probability  $\frac{1}{2}$ we find the overlap value $q =0$, which happens if in step 1 one ``odd'' and one ``even'' sequence was chosen,  that is, 
$k_1 - k_1'$ is odd.  
With probability $\frac{1}{8}$ one finds overlap values $q$ equal to  
plus or minus $\frac{1}{2}$ respectively, 
which happens if in the first step either the same 
($k_1=k_1'$), or the opposite ($|k_1- k_1'|=2$) choice was made, and in step 2 a
 different even-odd choice for $k_2$ and $k_2'$. We repeat this argument, and 
we find that with probability $\frac{1}{2^{2n+1}}$ one finds overlap values 
$\frac{(2p+1)}{2^{n}}$, with $p= -2^{n-1}, ......, 2^{n-1}-1$ \cite{Grothe}. 
Thus in the end, the overlap distribution becomes concentrated on the dyadic 
rationals.
\smallskip
\newline
Summarizing we have proven:

\begin{theorem}\label{theorempaperfolding}
 The overlap distribution of paperfolding sequences is given by:
\begin{equation}
 p(q) = \sum_{n=0}^{\infty} \sum^{'}_{m} \frac{1}{2} \left(\frac{1}{4} \right) ^{n} 
        \delta\left(q, \frac{m}{2^{n}} \right) 
\end{equation}
where the  integers  $n,m$  must satisfy the following  conditions:
\begin{itemize}
\item if $n=0$ then $m=0$,
\item if $n>0$ then $m$ is odd and $|m|<2^{n}$.
\end{itemize}
\end{theorem}

{\bf Comments:}

Properties which are common between the period-doubling and the 
paperfolding system are:
 
1) Limit periodicity. The sequences are a countable union of periodic 
subsequences of pluses and minuses. This implies a countable number of 
possible values for the overlap.

2) There is a tree (hierarchical, ultrametric) structure in the set of 
sequences-- the pure states--. 
As the overlap between two sequences
depends on the level of the tree at which a different even-odd choice is made, 
the overlap distribution reflects this.

3) Both paperfolding and period-doubling sequences form  model sets, 
obtained by a cut and project scheme, with an 
internal space of 2-adic numbers. \cite{BaMoSc,BaMoRicSi, Sc}. This agrees with 
the existence of a connection between Parisi's replica-symmetry-breaking ideas 
and p-adic numbers as was suggested in \cite{ParSou}.

\smallskip

However, in contrast to the period-doubling case, the set of paperfolding 
sequences is spin-flip symmetric, and the set on which the overlap 
distribution is concentrated now lies dense in the interval $[-1, +1]$.

\bigskip

\noindent {\em Acknowledgements}:
A.C.D. v.E  became interested in these issues during 
an earlier collaboration with Bert Hof and Jacek Mi\c{e}kisz.
He also acknowledges very useful discussions with Michael Baake, 
Jacek Mi\c{e}kisz  and Dimitri Petritis. We also thank Michael Baake for 
some helpful advice on the manuscript.

\end{document}